\documentclass[journal]{IEEEtran}
\ifCLASSINFOpdf
\else
\fi
\usepackage{amsmath}
\usepackage{dsfont}
\usepackage{amsfonts}
\usepackage{mathrsfs}
\usepackage{amsthm}
\usepackage{amssymb}
\usepackage{breqn}
\usepackage{graphicx}
\newlength{\dhatheight}

\usepackage{hyperref}
\usepackage{array}

\newtheorem{theorem}{Theorem}

\newtheorem{lemma}{Lemma}
\newtheorem{corollary}{Corollary}

\begin{document}
%\title{Novel Bounds on Finite-Length Hypothesis Testing Subject to Vanishing Type I Error Probabilities}
\title{ Finite-Length Bounds on Hypothesis Testing Subject to Vanishing Type I Error Restrictions}

\author{Sebastian Espinosa,
        Jorge F. Silva$^*$,
        and Pablo Piantanida,% <-this % stops a space
\thanks{S. Espinosa and J. F. Silva are with the Information and Decision System Group, Department of Electrical Engineering,  Universidad de Chile, Santiago, Chile (email: sebastian.espinosa@ing.uchile.cl, josilva@ing.uchile.cl).}
\thanks{This work of S. Espinosa was funded by the National Agency for Research and Development (ANID)/ScholarshipProgram/DoctoradoNacional/2018-21180693.}
\thanks{This project has received funding from the European Union’s Horizon 2020 research and innovation programme under the Marie Skłodowska-Curie grant agreement No 792464.}
\thanks{This article has been accepted for publication by IEEE Signal Processing Letters 2021  \cite{espinosa2021finite}. Personal use of this material is permitted.
  Permission from IEEE must be obtained for all other uses, in any current or future 
  media, including reprinting/republishing this material for advertising or promotional 
  purposes, creating new collective works, for resale or redistribution to servers or 
  lists, or reuse of any copyrighted component of this work in other works. 
  DOI: \href{<http://tex.stackexchange.com>}{10.1109/LSP.2021.3050381} }
%\thanks{P. Piantanida is with the Laboratoire des Signaux et Syst\`emes (L2S), CentraleSup\'elec-CNRS-Universit\'e Paris-Sud, France. % and with the Montreal Institute for Learning Algorithms (Mila), Universit\'e de Montr\'eal, Canada  
%(email: pablo.piantanida@centralesupelec.fr) The work of J.F. Silva was supported by Fondecyt 1170854 CONICYT-Chile and the Advanced Center for Electrical and Electronic Engineering, Basal Project FB0008. The work of Prof. Pablo Piantanida was supported by the European Commission’s Marie Sklodowska-Curie Actions (MSCA), through the Marie Sklodowska-Curie IF (H2020-MSCAIF-2017-EF-797805).}
}

\maketitle

\begin{abstract}
A central problem in Binary Hypothesis Testing (BHT) is to determine the optimal tradeoff between the Type I error (referred to as \emph{false alarm}) and Type II (referred to as \emph{miss}) error.  In this context, the exponential rate of convergence of the optimal miss error probability  --- as the sample size tends to infinity --- given some (positive) restrictions on the false alarm probabilities is a fundamental question to address in theory. Considering the more realistic context of a BHT with a finite number of observations, this paper presents a new non-asymptotic result for the scenario with monotonic (sub-exponential decreasing) restriction on the Type I error probability, which extends the result presented by Strassen in 2009. Building on the use of concentration inequalities, we offer new upper and lower bounds to the optimal Type II error probability for the case of  finite observations. 
%To illustrate their use, the derived bounds are evaluated and interpreted numerically for some vanishing Type I error restrictions and as a function of the number samples. 
Finally, the derived bounds are evaluated and interpreted numerically (as a function of the number samples) for some vanishing Type I error restrictions. 
\end{abstract}

\begin{IEEEkeywords}
Hypothesis testing, performance bounds,  finite-length analysis, error exponent, %feasibility, unfeasibility, 
concentration inequalities.
\end{IEEEkeywords}

% \fi
\IEEEpeerreviewmaketitle

\section{Introduction}
\label{sec_main}
Binary Hypothesis Testing (BHT) is a common problem in statistics and it has been richly used as a method  to statistical signal detection~\cite{4205085, 985674}.   In particular, the celebrated {\em Neyman-Pearson lemma} provides the optimal detection scheme for this testing task~\cite{4102537}.  
On the specifics, let us consider the classical $n$-length BHT setting given by 
%------------------------------------------------------------------------------------%
\begin{equation}
\left\{\begin{array}{lll}
H_0:  &\ X_1^n \sim  P^n,\nonumber\\
%\theta=1\rightarrow 
H_1: &\ X_1^n \sim Q^n,
\end{array} \right. \label{eq_sec_main_1}
\end{equation}
%------------------------------------------------------------------------------------%
where $P, Q \in \mathcal{P}(\mathbb{X})$ with  $D(P\|Q)>0$. In this work, we restrict our attention to the case of a finite-alphabet $\mathbb{X}$, where $\mathcal{P}(\mathbb{X})$ denotes the family of probabilities on $\mathbb{X}$.   A decision rule $\phi_n$ of length $n$ is a function $\phi_n: \mathbb{X}^n \rightarrow \Theta \triangleq  \{0,1 \}, $
%------------------------------------------------------------------------------------%
%\begin{equation*}
%\phi_n: \mathbb{X}^n \rightarrow \Theta \triangleq  \{0,1 \}, 
%\end{equation*}
%------------------------------------------------------------------------------------%
%For any $\phi_n$ of length $n$ and distribution $P, Q$,
from which two types of errors are induced \cite{kullback1951information}:
%------------------------------------------------------------------------------------%
$$
\begin{array}{lll}
%\label{eq_sec_main_2}
P_0(\phi_n)& \equiv & P^n( \left\{ x_1^n \in \mathbb{X}^n: \phi_n(x_1^n) \neq 0 \right\}) \triangleq P^n(\mathcal{A}^c(\phi_n)),\\
P_1(\phi_n)& \equiv &  Q^n(  \left\{ x_1^n\in \mathbb{X}^n:  \phi_n(x_1^n) = 0 \right\} ) \triangleq Q^n(\mathcal{A}(\phi_n) ),%\label{eq_sec_main_3}
\end{array}
$$
with  decision region $\mathcal{A}(\phi_n) \triangleq \{ x_1^n\in \mathbb{X}^n : \phi_n(x_1^n) = 0\}$. 

%In this paper, we investigate a specific family of optimal decisions. 
For a given sequence $(\epsilon_n)_{n}$ of non-negative values such that $\lim_{n\rightarrow \infty}\limits \epsilon_n=0$, we study the solution to: 
%------------------------------------------------------------------------------------%
\begin{equation}
\beta_n(\epsilon_n) \equiv \min_{\begin{subarray}{c} \phi_n \in \Phi_n 
%\text{ operating at type I error }\epsilon_n \\
\end{subarray}} \{P_1(\phi_n): \text{ s.t. }  P_0(\phi_n)\leq \epsilon_n \},  \forall n\geq 1, 
\label{typeII}
\end{equation}
where $\Phi_n \equiv  \left\{\phi_n: \mathbb{X}^n \rightarrow  \Theta \right\}$ denotes the class of $n$-length detectors.
Importantly, $(\beta_n(\epsilon_n))_{n\geq 1}$ represents the optimum Type II error sequence that satisfies a sequence of fixed Type I error constraints. 

The \emph{Neyman-Pearson} lemma~\cite{neyman1933ix} offers the optimal trade-off between the two type of errors\footnote{See \cite{6210366} for a new proof based on properties of exponential  density function families.}. In this context, the determination of the (exponential) rate of convergence of the Type II error, which is known as the error exponent, has been a central problem in HT's analysis. Indeed, the error exponent is seen as an indicator of the complexity of the decision task (function of $P_0$, $P_1$ and $(\epsilon_n)_{n}$) and has found numerous applications \cite{marano_2019,Wang_2018}. For the important case when $\epsilon_n=\epsilon>0$ for all $n$, the celebrated {\em Stein's lemma} establishes that the error exponent of the Type II error is given by the KL divergence  $D(P\|Q) \equiv \sum_{x\in \mathbb{X}} P(x) \log \frac{P(x)}{Q(x)}$ \cite{kullback1951information,cover2012elements}. %It follows that:  
%------------------------------------------
\begin{lemma}[\textit{Stein's lemma}~\cite{chernoff1952measure,cover2012elements}]
	\label{lemma_stein}
	%Under the setting presented in Eq.(\ref{eq_sec_main_1}), 
	For any fixed $\epsilon \in (0,1)$, %it follows that
	%\begin{align*}%\label{eq_intro_3}
	$\lim_{n \rightarrow \infty} -\frac{1}{n } \log (\beta_n(\epsilon)) =  D(P\|Q)$.
	%\end{align*}
\end{lemma}
Importantly, % in this context, 
the error exponent limit in Lemma \ref{lemma_stein} is independent of $\epsilon>0$.  
However, this limit changes when we impose a setting with a monotonic decreasing Type I error restrictions.
%as a function of the number of observations. 
In particular, Han {\it et al.} \cite{han1989exponential} studied the case when the Type I error sequence has an exponential decreasing behaviour. Nagakawa {\it et al.} \cite{nakagawa1993converse} extended this analysis for a %more general 
family of decreasing sequence of Type I error restrictions: %This result is the following:  
%==================
\begin{lemma}\cite[Nakagawa]{nakagawa1993converse}
\label{nakagawalemma}
Let us assume that $\epsilon_n \leq e^{-rn}$ for some $r \in (0,D(P\|Q))$, then
%\begin{equation} 
$\lim_{n\rightarrow \infty}-\frac{1}{n} \log (\beta_n(\epsilon_n))=D(P_{t^{\ast}}\|Q)$,
%\label{Nakagawa}
%\end{equation}
where %$P_{t^{\ast}}$ denotes 
$P_{t^{\ast}}(x)\equiv  C_{t^{\ast}}P(x)^{1-t^{\ast}}Q(x)^{t^{\ast}}$ $\forall x \in \mathbb{X}$, and $t^{\ast}$ is the solution of $D(P_{t^\ast}\|P)=r$.
\end{lemma}
A direct implication of Lemma \ref{nakagawalemma} is the following result: 
%\dots\dots\dots\dots\dots\dots\dots\dots\dots\dots\dots\dots\dots\dots.. 
\begin{corollary}\label{cor_nakagawalemma} \cite{nakagawa1993converse} %\cite[Sect. IX]{nakagawa1993converse}
Let us assume that %${1}/{\epsilon_n} \triangleq o(e^{n})$,
            %$\forall r >0,$ 
            $({1}/{\epsilon_n})_n$ is  $o(e^{rn})$ for any $r>0$,
 then %\cite[Sect. IX]{nakagawa1993converse}
%\begin{equation} 
$\lim_{n\rightarrow \infty} - \frac{1}{n} \log (\beta_n(\epsilon_n))=D(P\|Q)$. 
%\label{Nakagawa2}
%\end{equation}
\end{corollary}
Importantly,  Corollary \ref{cor_nakagawalemma} shows that the same %closed-form expression for 
%the 
error exponent of the Stein's lemma is obtained for these stringent family of %decision 
problems --- where $(\epsilon_n)_{n}$ tends to zero at a sub-exponential rate. In contrast, when the Type I error restriction tends to zero exponentially fast (Lemma \ref{nakagawalemma}), the error exponent is strictly smaller than $D(P\|Q)$.   
%from (\ref{Nakagawa}), 
%yielding a discrepancy with respect to the less restrictive regime covered by the {\em Stein's} lemma \cite{chernoff1952measure}.

%\subsection{Finite-length Analysis of Hypothesis Testing}
%___________________________________________________
%___________________________________________________
\subsection{Finite-Length Context  and Contribution}
\label{sub_sec_unconstrained_finite_length_result} 
In many practical problems, the statistician has access only to a finite number of observations. Consequently,  it is critical to obtain non-asymptotic bounds for the probability of error $\beta_n(\epsilon_n)$ for a finite $n$. 
%Therefore, it is relevant not only to know the error exponent expression but also to derive finite-length performance bounds for $\beta_n(\epsilon_n)$. 
%
Concerning the non-asymptotic analysis of this problem, the following  result was derived by Strassen for the specific regime when $\epsilon_n=\epsilon>0$ for all $n\geq 1$ \cite{strassen2009asymptotic}.
%====================================
\begin{lemma} \label{th_strassen2009asymptotic} \cite{strassen2009asymptotic} 
Let us consider $\epsilon \in (0,1)$, then eventually with $n$, it follows that
%---------------------------------------------s---------------------------------------%
%\begin{dmath*}
$-\frac{\log(\beta_n(\epsilon))}{n}=D(P\|Q)+\sqrt{\frac{V(P\|Q)}{n}}\Phi^{-1}(\epsilon)+\frac{\log n}{2n} + \mathcal{O}\left (\frac{1}{n} \right)$,
%\end{dmath*}
where $V(P\|Q) \equiv \displaystyle \sum_{x \in \mathbb{X}}\limits P(\{x\})\left [\log \left (\frac{P(\{x\})}{Q(\{x\})} \right ) -D(P\|Q) \right ]^2$.
\end{lemma} 
Lemma \ref{th_strassen2009asymptotic} %offers a non-asymptotic expression of the form: 
shows that 
%------------------------------------------------------------------------------------% 
%\begin{corollary} %(Second order of a probability of type II error) 
%\label{cor_stressen_rate_conv}
%Eventually in $n$ it follows that
%\begin{equation} 
$\left \rvert D(P\|Q)- \left( -\frac{1}{n}\log(\beta_n(\epsilon)) \right )\right \rvert$ is $\mathcal{O}\left(\frac{1}{\sqrt{n}}\right)$,
%\label{overheadNR}
%\end{equation} 
%\end{corollary}
which expresses the velocity of convergence of $ -\frac{1}{n}\log(\beta_n(\epsilon))$ %(the non-asymptotic error exponent) 
to its %asymptotic
 limit $D(P\|Q)$.
Given the practical importance of this type of finite length results, % for applications that have a finite number of observations, 
it is very relevant to derive new results that extend  %the context where 
Lemma \ref{th_strassen2009asymptotic} %applies in 
to our general problem in (\ref{typeII}),  
as a function of $P$, $Q$,  $(\epsilon_n)_n$ and  $n$. In addition, it is critical that these bounds can be evaluated for its practical use. This last aspect is not achieved in Lemma \ref{th_strassen2009asymptotic}, which from that perspective is an asymptotic (convergence) result.

The main contribution of this paper goes in this direction, where we derive new upper and lower bounds for the discrepancy between $- \frac{1}{n}\log (\beta_n(\epsilon_n))$ and its information limit $D(P\|Q)$ for any finite $n\geq 1$ when $(\epsilon_n)_{n}$ tends to zero at a sub-exponential rate. 
These expressions can be evaluated and interpreted numerically in any context where we know the models ($P$ and $Q$) and the parameters of the problem ($\epsilon_n$ and $n$). In addition, these new bounds stipulate the velocity at which the error exponent is achieved as the sample size tends to infinity. From this, we could assess how realistic the information limits (asymptotic results) are in practice when facing a problem with a finite number of observations. To conclude our analysis, we numerically compute and evaluate the expressions obtained by our result to show the derived bounds' tightness for some specific scenarios.

\subsection{Related Work} 
In a Bayesian setting, % of this problem, 
Sason \cite{sason2012moderate} obtained an upper bound to the optimal Bayesian probability of error (non-asymptotic) by bounding the Type I and Type II errors simultaneously in such a way that they both decay to zero sub-exponentially with $n$. 
%NOT REALLY RELEVANT.......
%This interesting result was derived from a {\em moderate deviation analysis} (MDA) approach \cite{zeitouni1998large}. 
%NOT REALLY RELEVANT.......--
%It should be pointed out that 
It is worth to mention that this work differs from the current setting in the sense that we are interested in bounding the discrepancy between $- \frac{1}{n}\log (\beta_n(\epsilon_n))$ and its information limit and how this analysis depends on the vanishing Type I error restrictions. In addition, we are interested in the velocity of convergence of  $- \frac{1}{n}\log (\beta_n(\epsilon_n))$ to its  information limit %in the large deviation regime 
and the impact of considering stringent restriction on Type I errors $(\epsilon_n)_n$. %t is worth to mention that 
%On addition, 
Complementing this paper, 
\cite{espinosa2019new} studies %a similar problem  but in 
a distributed (two-terminal) version of the BHT problem subject to communication (rates) constraints. Our results here  do not derive from \cite{espinosa2019new} since the setups are very different from each other, and different tools are used to address them. 
Finally,  a similar %error exponent 
analysis of the Type I error has been addressed by Bahadur \cite{nikitin1995asymptotic}. In contrast to this work's focus, this analysis considers  %classical Neyman-Pearson approach, 
a fixed restriction on the power of a test ($1-$Type II error) to determine the exponential rate of convergence of their sizes (Type I error) as $n$ %the number of observations 
tends to infinity.
%More precisely, there is an inverse proportional relationship between the error exponent of the false alarm probability with the number of observations for a prescribed power of test. [REDUNDANT]

%More precisely, in \cite{espinosa2019finite} we only have access to an encoded version of the samples. This scenario introduces new technical difficulties and is a more challenging problem. Therefore, the tools presented in this work cannot be applied directly.

\subsection{Notations and Organization}
%We restrict our attention to the case of a finite-alphabet space $\mathbb{X}$, where $\mathcal{P}(\mathbb{X})$ denotes the family of probabilities on $\mathbb{X}$. 
%NON REALLY NEEDED FOR SUCH AS SHORT PAPER --------------------
%Boldface letters ${x_1^n}$ and upper-case letters $X_1^n$ are used to denote vectors and random vectors of length $n$, respectively. 
%NON REALLY NEEDED FOR SUCH AS SHORT PAPER --------------------
%Let $(b_n)_n \equiv o (a_n)$ 
$(b_n)_n$ being $o(a_n)$ indicates that $\limsup_{n\rightarrow \infty} \left(b_n/a_n\right)=0$ and %$(b_n)_n\equiv  \mathcal{O} (a_n)$ 
$(b_n)_n$ being  $\mathcal{O} (a_n)$ 
indicates that $\limsup_{n\rightarrow \infty}  | b_n/a_n|  <\infty $. We say that $(f(n))_n \approx (g(n))_n$ if  there exists a constant $C>0$ such that $f(n)=Cg(n)$ eventually in $n$. %, for all $n\geq N$.  
The rest of the paper is organized as follows: Section \ref{sec_finite_lenght} presents the main result of this work. % for the non-asymptotic regime. 
Numerical analysis and discussions are presented in Section \ref{numan}. The proof of  is in Sect.~\ref{proofresult3}.
 
%================================================
\section{Main Result}
\label{sec_finite_lenght}
The main result of this letter extends Lemma \ref{th_strassen2009asymptotic} offering new non-asymptotic bounds for $\beta_n(\epsilon_n)$ in (\ref{typeII}) under sub-exponential Type I error restrictions.  %Corollary \ref{cor_nakagawalemma}. 
In particular, the next result provides upper and lower bounds for the discrepancy between $- \frac{1}{n}\log (\beta_n(\epsilon_n))$ and $D(P\|Q)$.  %and from this analysis the finite-length value of $-\frac{1}{n}\log (\beta_n(\epsilon_n))$ as $n$ tends to infinity. The following theorem offers that extension.
%------------------------------------------------------------------------------------%
\begin{theorem}[]\label{theorem3}
Let us assume that $P\ll Q$  and  %$(1/\epsilon_n) \equiv o(e^{n})$.
                                 %$\forall r >0,$ 
                                 that $(1/\epsilon_n)_n$ is  $o(e^{rn})$ for any $r>0$. 
Then, eventually in $n$, it follows that:
$$
\begin{array}{ll}
 -\frac{1}{n}\log(\beta_n(\epsilon_n)) \geq   D(P\|Q) - \displaystyle C_X(P,Q) \sqrt{\frac{2\ln(1/\epsilon_n)}{n}} \\
-\frac{1}{n}\log(\beta_n(\epsilon_n)) \leq  D(P\|Q)+\frac{\displaystyle  \log\left (\displaystyle \frac{1}{1-\epsilon_n-\delta_{n}}\right )}{\displaystyle  n} +\delta_{n}
\end{array}
$$
%$+\delta_{n}$ 
where $C_X(P,Q) \equiv  \sup_{x \in \mathbb{X}}\limits \left \rvert \log\left( \frac{P(\{x \})}{Q(\{ x\})}\right)\right \rvert$ and $\delta_{n} \equiv  C_X(P,Q) \sqrt{\frac{2\ln(1/\epsilon_n)}{n}}$.
\end{theorem}
%The proof is presented in Section \ref{proofresult3}.

\subsection{Interpretation and Discussion of Theorem \ref{theorem3}}
%\begin{itemize}
%\item 
{\bf 1:} This result establishes a non-asymptotic rate of convergence for the Type II error when we impose a vanishing condition on  $(\epsilon_n)_{n}$ that is sub-exponential. Interestingly, the bounds for the discrepancy $-\frac{1}{n}\log(\beta_n(\epsilon_n))$ % obtained as a function of $n$
depend explicitly on the sequence $(\epsilon_n)_{n}$.\\  
%
%\item 
{\bf 2:} It is worth noting that the dependency on $(\epsilon_n)_{n}$ observed in our result is non-observed in the asymptotic limit in Corollary \ref{cor_nakagawalemma},  %Lemma \ref{Nakagawa}, 
which is  $D(P\|Q)$ as long as $(1/\epsilon_n)_{n}$ is sub-exponential.\\ % with $n$. 
%
%\item 
{\bf 3:} Adding on the previous point, the fact that the asymptotic error exponent is invariant from the simpler fixed Type I setup  (in Lemma \ref{lemma_stein}) %in Eq.(\ref{Nakagawa2}) 
to the more restrictive sub-exponential Type I error decay setting (in Corollary \ref{cor_nakagawalemma}), it is however manifested in our non-asymptotic result  in term of the rate of convergence to the limit $D(P\|Q)$. % with respect to the result in Lemma \ref{th_strassen2009asymptotic}. 
In particular,  there is a concrete penalty  $\mathcal{O}(\sqrt{\log(1/\epsilon_n)})$ on the velocity of convergence to zero of the discrepancy $( -\frac{1}{n}\log \beta_n(\epsilon_n) - D(P\|Q))$  in our result compared with what is obtained in Lemma \ref{th_strassen2009asymptotic}.\\ %(\ref{overheadNR}).
%
%\item 
{\bf 4:} The proof of the Theorem~\ref{theorem3} has two parts: the constructive and unfeasibility arguments. Both arguments are constructed from concentration inequalities using the i.i.d. structure of the observations. For the constructive argument,  we apply the bounded difference inequality \cite{boucheron2013concentration}. On the unfeasibility argument,  we use (concentration) results from typical sequences \cite{cover2012elements} to construct a lower bound on the minimum probability of Type II error.\\ 
%JORGE: This is unclear to me.....
%Interestingly, the result on the unfeasibility argument is higher order accurate compared to the constructive case.
%
%\item 
{\bf 5:} %It is worth noting that 
If we impose a fixed value of $\epsilon_n=\epsilon \in (0,1)$,  our result recovers the rate of convergence for the Type II error given by Lemma \ref{th_strassen2009asymptotic}. However, we obtained explicit bounds.\\
%OMITED: THIS REMARK IS NOT CENTRAL FOR OUR RESULT -------------------------------------
%\item 
%{\bf 6:} Finally, when we use an exponential rate $\epsilon_n\equiv O(e^{-n})$, the bounds presented in our argument do not converge to zero. This is consistent with  the fact that the fundamental limit of the BHT problem changes when $\epsilon_n\equiv \mathcal{O}(e^{-n})$,  as seen in Lemma \ref{nakagawalemma}.
%\end{itemize}
%-----------------------------------------------------------------------------------------------------------------------------

%================================
%================================
\section{Practical Implications of Theorem~\ref{theorem3}}
\label{numan}
In this section, we show how Theorem \ref{theorem3} may be adopted by a statistician to obtain  bounds on  $\beta_n(\epsilon_n)$ when $n$ is finite. The resulting bounds %in Theorem \ref{theorem3} 
provide an interval of feasibility for %the optimal Type II error probability 
$\beta_n(\epsilon_n)$:
\begin{align*}
 \textrm{UB}(\epsilon_n) & \equiv \exp\Big [-n\Big (D(P\|Q) - \sqrt{\frac{2\ln(1/\epsilon_n)}{n}}C_X(P,Q)\Big ) \Big ], \\
 \textrm{LB}(\epsilon_n) & \equiv %\nonumber \\
  \exp\Big [-n\Big (D(P\|Q)-\frac{1}{n}\log\big (1-\epsilon_n-\delta_{n}(\epsilon_n)\Big ) \nonumber \\
& +\delta_{n}(\epsilon_n) \big ) \Big ].
%\label{Converse_bound}
\end{align*}
The length of $[\textrm{LB}(\epsilon_n),\textrm{UB}(\epsilon_n)]$ indicates the precision of the result %of the error probability %by its exponent,
and, at the same time, the interval $[\textrm{LB}(\epsilon_n),\textrm{UB}(\epsilon_n)]$  
%measures the quality of this approximation, i.e., the gap between $\beta_n(\epsilon_n)$ and $e^{-nD(P\|Q)}$.
can be used to measure how close $\beta_n(\epsilon_n)$ is to $e^{-nD(P\|Q)}$.
%=================================================================
%=================================================================
\begin{figure}
    \centering
    \begin{minipage}{0.5\textwidth}
        \centering
        \includegraphics[width=1.0\linewidth]{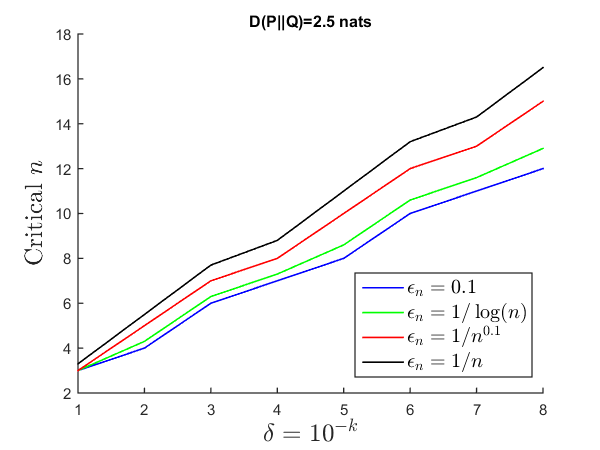} 
        \caption{Critical number of samples (CSS) predicted by Th. \ref{theorem3}  %$n^{\ast}(\epsilon_n,\delta)$ 
        across different values of $\delta=10^{-k}$. %, with $k \in \{1,...,8 \}$.  
        High divergence case with $D(P\|Q)=2.5$ and  $C_X(P,Q)=2.04$. % nats
        }
        \label{High}
    \end{minipage}\hfill
    \begin{minipage}{0.5\textwidth}
        \centering
       \includegraphics[width=1.0\linewidth]{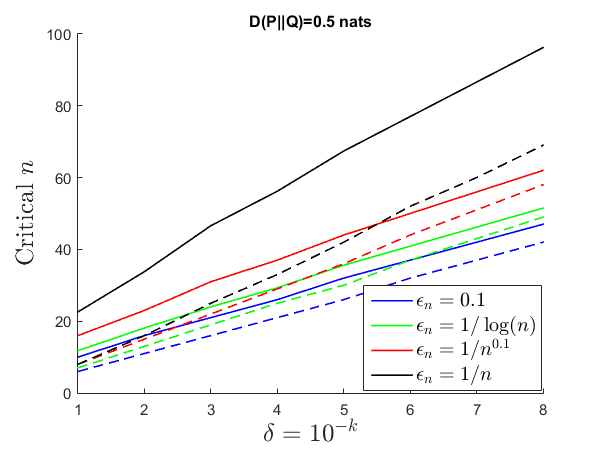} 
        \caption{%Critical number of samples 
        CSS predicted by Th. \ref{theorem3} % $n^{\ast}(\epsilon_n,\delta)$ 
        across different values of $\delta=10^{-k}$. %, with $k \in \{1,...,8 \}$. 
        Low divergence case with $D(P\|Q)=0.5$  and  $C_X(P,Q)=1.03$. % nats. 
        The dashed lines show an estimation of  the exact  CSS obtained from $\beta_n(\epsilon_n)$ directly. 
        %for the same velocities for $(\epsilon_n)_{n\geq 1}$
        }
           \label{Low}
    \end{minipage}
\end{figure}
%=================================================================
%=================================================================

\begin{table*}[ht]
\centering
\begin{tabular}{|c|c|c|c|c|c|c|c|c|}
\hline
  & \multicolumn{8}{c|}{Number of observations $n$ } \\
\hline
 $\epsilon_n$     & 50 & 150 & 250 & 350 & 450 & 550 & 650 & 750  \\
\hline
 $1/\log(n)$  &     2.3587e-10 & 3.3962e-45 & 1.0595e-83& 9.4592e-124 & 1.0229e-164& 2.6103e-206& 2.2862e-248 & 8.6970e-291 \\
$1/n^{0.1}$  &  7.8229e-17 &  8.1724e-57 &  9.1096e-99& 1.3994e-141& 6.4863e-185&  1.3117e-228 & 1.4272e-272&  9.5440e-317\\
$1/n$   &       0.5571  & 3.7757e-25& 7.4403e-56& 2.7823e-89& 2.3527e-124& 1.4443e-160& 1.2489e-197& 2.3163e-235 \\
\hline
\end{tabular}
\vspace{2mm}
\caption{Magnitude of $  \textrm{UB}(\epsilon_n)- \textrm{LB}(\epsilon_n)$ function of $\epsilon_n$ and $n$ for the case when $D(P\|Q)=1$.}
\label{Tablap}
\end{table*}

Table \ref{Tablap} presents the length of $[\textrm{LB}(\epsilon_n),\textrm{UB}(\epsilon_n)]$ for three regimes of:  $\epsilon_n \in \{n^{-1},n^{-0.1},1/\log(n) \}$, and  two models $P$, $Q$ where  $D(P\|Q)=1$ with $|\mathbb{X}|=15$. First, we observe that the length of $[\textrm{LB}(\epsilon_n), \textrm{UB}(\epsilon_n)]$ vanishes exponentially fast 
with the sample size. From this exponential decay, we observe that the centered value predicted by Theorem \ref{theorem3}, i.e., the exponential behavior  $\exp(-n D(P\|Q))$,  is a good approximation for $\beta_n(\epsilon_n)$ provided that $n$ is sufficiently large. This  supports the idea that 
 $\exp(-n D(P\|Q))$ is a useful proxy for  $\beta_n(\epsilon_n)$ provided that a Critical Sample Size (CSS) is achieved (more details on this below).
Table \ref{Tablap} also shows that the result's precision is affected by the velocity of convergence of the Type I error restriction $(\epsilon_n)_{n}$, which is consistent with the statement and the analysis of our main result. In particular, for a faster speed of convergence of $(\epsilon_n)_{n}$ to zero (i.e., a stringer problem), the gap between the bounds is more prominent, which means that the bounds of Theorem \ref{theorem3} are expected to be less informative about  $\beta_n(\epsilon_n)$. 

%for a very small number of observations. 
%REVIEWER CONCERN ON THE PRACTICALITY OF THIS STATEMENT: ---------------------(we omitted this statement that is confusing)
%This issue can be attributed partially to the value of $C_X(P,Q)$, representing a conservative worse-case constant in our analysis. Nevertheless, after a reasonable sample-size, the effect of this constant vanishes very quickly since our bounds are exponentially decreasing with $n$.

Regarding  the implications of the above bounds to  measure the gap between $\beta_n(\epsilon_n)$ and  $e^{-nD(P\|Q)}$, we address the
following question: given an arbitrary value of $\delta>0$ of the form $10^{-k}$ with $k \in \left\{1,\dots ,8 \right\}$, and for two arbitrary models $P$ and $Q$, we want to predict from Theorem \ref{theorem3} the minimum number of samples required  to guarantee that $\beta_n(\epsilon_n) \in (e^{-nD(P\|Q)}- \delta, e^{-nD(P\|Q)} + \delta)$. The exponential decay of the length of $[\textrm{LB}(\epsilon_n),\textrm{UB}(\epsilon_n)]$,  observed in Table  \ref{Tablap}, implies that this should  happen eventually with $n$ very quickly. Indeed, we can derive an upper bound for this critical number of samples (CSS) from the expressions we have for $\textrm{LB}(\epsilon_n)$ and $\textrm{UB}(\epsilon_n)$.\footnote{The predicted CSS is the first $n\geq 1$ such that $\max\{ \textrm{UB}(\epsilon_n)- e^{-nD(P\|Q)},e^{-nD(P\|Q)}-\textrm{LB}(\epsilon_n) \}  \leq \delta$, which is finite for any $\delta>0$.}  Figures \ref{High} and \ref{Low} present the predicted CSS versus  $\delta= 10^{-k}$ for different scenarios of $P$, $Q$ (in terms of $D(P\|Q)$) and $(\epsilon_n)_n$. We consider two scenarios for $P$ and $Q$ (low divergence $D(P\|Q)=0.5$ and high divergence $D(P\|Q)=2.5$) and we explore $(\epsilon_n)_n  \in \{n^{-1}, n^{-0.1},1/\log(n), 0.1\}$. Figures \ref{High} and \ref{Low} show that even for really small precision $\delta=10^{-8}$ the point at which $\beta_n(\epsilon_n)$ can be well approximated by $e^{-nD(P\|Q)}$ requires at most $16$ samples  and  $60$ samples for high and low divergence cases, respectively, and   the majority of $(\epsilon_n)_n$. The dependency of these curves on the magnitude of $D(P\|Q)$ and $(\epsilon_n)_n$ is clearly expressed in these findings, which is consistent with our previous analyses.

Finally, to evaluate the tightness of our predictions, we simulate i.i.d. samples according to $P$ and $Q$ from which a  precise empirical estimation of $\beta_n(\epsilon_n)$ is derived. In particular, given $P$, $Q$ and $(\epsilon_n)_n$, we obtained empirical estimations of the error probabilities (Type I and Type II) from which we estimate $\beta_n(\epsilon_n)$. For this purpose,  $2.5\cdot 10^6$ realizations of $P$ and $Q$ were used to have good estimations of these probabilities.  Using the estimated values of  $\beta_n(\epsilon_n)$, we obtain the point where $\beta_n(\epsilon_n) \in (e^{-nD(P\|Q)}- \delta, e^{-nD(P\|Q)} + \delta)$ directly.  Figure \ref{Low} contrasts our predictions and the true (estimated) values (the dashed lines) of the CSS. 
%obtained from the empirical estimation of $\beta_n(\epsilon_n)$.
Consistent with our result's nature, our prediction of the CSS is more conservative than the true CSS estimated from simulations. % ( from the true (estimated) $\beta_n(\epsilon_n)$).  
Importantly, this discrepancy is not significant overall, expressing that our bounds are useful for this analysis and can be adopted in cases where it is impractical to estimate $\beta_n(\epsilon_n)$ from data. Indeed, in this analysis, we face this issue, and it is very difficult to obtain accurate estimates of  $\beta_n(\epsilon_n)$ for high divergence regimes. Notice that $\beta_n(\epsilon_n)$ is of order:  $O(e^{- n D(P\|Q)})$ for which  around $e^{n D(P\|Q)}$ simulations (i.e., i.i.d. samples from $P$ and $Q$) are needed. This becomes  impractical even for $n$ less than $30$ when $D(P\|Q)$ is relatively large.

\section{Proof of Theorem \ref{theorem3}}
\label{proofresult3}
We divide the proof of Theorem \ref{theorem3} in two parts.  %lower and upper bound analysis.
%-----------------------------------------------------------------------------------%

\subsubsection{Lower Bound Analysis}
%\begin{lemma}
Under the assumption of Theorem \ref{theorem3},  let us verify that %$(1/\epsilon_n) \equiv o(e^{n})$, then
            %        $\forall r >0,$ $(1/\epsilon_n)_n$ is $o(e^{rn})$, then
\begin{equation*}
 D(P\|Q)- \left( -\frac{1}{n}\log \beta_n(\epsilon_n) \right )\leq \sqrt{\frac{2\ln(1/\epsilon_n)}{n}}C_X(P,Q).
\end{equation*}
%\end{lemma}
%-----------------------------------------------------------------------------------%
%\begin{proof}
Let us consider the corresponding optimal decision regions from the Neyman-Pearson Lemma parameterized in the following way: $\forall t >0$,
\begin{equation}
\mathcal{B}_{n,t}= \left \{x_1^n \in \mathbb{X}^n : \frac{P^n(\{x_1^n \})}{Q^n(\{x_1^n \})} > e^{nt}\right \}.
\end{equation}
Considering the induced test $\phi_{n,t}(\cdot): \mathbb{X}^n \mapsto \{0,1 \}$ such that $\phi_{n,t}^{-1}(\left\{ 0 \right\})=\mathcal{B}_{n,t}$. The Type I error probability is given by $P^n(\mathcal{B}^c_{n,t})$. An upper bound for the Type II follows as:
\begin{equation}
Q^n\left (\mathcal{B}_{n,t}\right )  \leq e^{-nt}.
\label{typeIIerrorcound}
\end{equation}
Then, for any finite $n>0$ and $\epsilon_n>0$,  finding an achievable Type II error exponent from this construction (and the bound in Eq.(\ref{typeIIerrorcound})) reduces to solve the following 
problem:
%------------------------------------------------------------------------------------%
\begin{equation}
t^{\ast}_n(\epsilon_n)\triangleq\sup_t \{t: % P_{X,Y}^n(B_{n,t}^c(f_n)) 
P^n (\mathcal{B}^c_{n,t}) \leq \epsilon_n \}.
\label{optimumt2}
\end{equation}
It will be convenient to re-parameterize $t$  with respect to 
the value $D(P\|Q)$. More precisely, let us define
%\dots\dots\dots
$$t_{\delta}\triangleq  D(P\|Q) -\delta,$$ for any $\delta >0$. Then using the bounded difference inequality \cite{boucheron2013concentration},  we obtain
\begin{align} \label{eq_lemma_achievability_th4}
P^n \left (	B^c_{n,t_{\delta}} \right )&=P^n \left (	x_1^n \in \mathbb{X}^n :   \Biggr \rvert \hat{D}(P\|Q)-D(P\|Q) \Biggr \rvert \geq \delta  \right ) \nonumber \\
& \leq \exp\left (  \frac{-n\delta^2}{2C_X(P,Q)^2}\right ),
\end{align}
where $\hat{D}(P\|Q) \triangleq \frac{1}{n} \sum_{i=1}^n \log \left (\frac{P(\{x_i\} )}{Q(\{x_i\})} \right )$ is the empirical divergence. Finally, from Eq. (\ref{optimumt2}) a lower bound for $t^{\ast}_n(\epsilon_n)$ can be determined from Eq. (\ref{eq_lemma_achievability_th4}) by letting  $\tilde{\delta}_{n}(\epsilon_n)$ to be the solution of the following equality:
%------------------------------------------------------------------------------------% 
\begin{equation}
\exp \left ( \frac{-n\tilde{\delta}_{n}(\epsilon_n)^2}{2C_X(P,Q)^2}\right )=\epsilon_n.
\end{equation}
%------------------------------------------------------------------------------------% 
Consequently, we have that
%------------------------------------------------------------------------------------% 
\begin{equation}
t^{\ast}_{n}(\epsilon_n)\geq  t_{\tilde{\delta}_{n}(\epsilon_n)}\triangleq D(P\|Q) -\sqrt{\frac{2\log(1/\epsilon_n)}{n}}C_X(P,Q).
\label{boundupperR2}
\end{equation}
Finally, replacing the bound of (\ref{boundupperR2}) in (\ref{typeIIerrorcound}) and taking logarithm we have that:
\begin{equation}
 D(P\|Q)- \left( -\frac{1}{n}\log(\beta_n(\epsilon_n)) \right )\leq \sqrt{\frac{2\ln(1/\epsilon_n)}{n}}C_X(P,Q), 
\end{equation}
which concludes this part. 
%\end{proof}

\subsubsection{Upper Bound Analysis}
%\begin{lemma}
%Let us assume that %$(1/\epsilon_n) \equiv o(e^{n})$ 
%                    $\forall r >0,$ $(1/\epsilon_n)_n$ is  $o(e^{rn})$ 
%then, eventually in $n$,
%\begin{equation}
%-\frac{1}{n}\log(\beta_n(\epsilon_n))  \leq D(P\|Q)+\frac{\log\left (\frac{1}{1-\epsilon_n-\delta_{n}(\epsilon_n)}\right )}{n}+\delta_{n}(\epsilon_n)
%\end{equation}
%with $\delta_n(\epsilon_n)=\sqrt{\frac{2\ln(1/\epsilon_n)}{n}}C_X(P,Q).$
%\end{lemma}
%\begin{proof}
Let us consider the set
%------------------------------------------------------------------------------------%
\begin{equation}
\mathcal{A}_{n,\delta }^c\triangleq \left \{ x_1^n\in \mathbb{X}^n : \Biggr \rvert \frac{1}{n}\log \left (\frac{P^n(\{ x_1^n \} )}{Q^n(\{ x_1^n \}  )} \right ) -  D(P\|Q) \Biggr \rvert \geq \delta  \right \},
\label{Adelta}
\end{equation}
for any $\delta>0$. 
%------------------------------------------------------------------------------------%
We have the following result:
%------------------------------------------------------------------------------------%
\begin{lemma}
 \cite[Sect 11.8]{cover2012elements} 
 \label{lemma_needed}
 For any set $ \mathcal{B}_n \subseteq \mathbb{X}^n$ and its induced test $\phi_n$\footnote{Meaning that $\phi_{n}(x_1^n)=0$ if $x_1^n\in \mathcal{B}_{n}$.} such that operates at Type I error $\epsilon_n$ (i.e. $P^n(	\mathcal{B}^c_n ) \leq \epsilon_n$), then
%------------------------------------------------------------------------------------%
\begin{equation}
Q^n(\mathcal{B}_n) \geq (1-\epsilon_n-\delta)2^{-n(D(P\|Q)+\delta)}.
\label{cover_converse}
\end{equation}
%------------------------------------------------------------------------------------%
\end{lemma} 
\noindent By construction, it is clear that there exists $\delta >0$ such that $\mathcal{A}^c_{n,\delta }$ operates at Type I error $\epsilon_n$. In fact, we consider
%------------------------------------------------------------------------------------%
\begin{equation}
\delta^{\ast}_{n} \triangleq \sup \{\delta: P^n(\mathcal{A}^c_{n,\delta }) \leq \epsilon_n \}.
\label{36delta}
\end{equation}
Using the bounded difference inequality \cite{boucheron2013concentration}, we get that
\begin{align} \label{eq_lemma_converseth4}
P^n \left (	\mathcal{A}_{n,\delta }^c \right )&=P^n \left (x_1^n \in \mathbb{X}^n :   \Biggr \rvert \hat{D}(P\|Q)-D(P\|Q) \Biggr \rvert \geq \delta \right )\nonumber \\
& \leq \exp\left (  \frac{-n\delta^2}{2C_X(P,Q)^2}\right ).
\end{align}
Using the same argument from the lower bound analysis, we obtain a lower bound for $\delta^{\ast}_{n}$ given by
\begin{equation}
\delta^{\ast}_{n} \geq \delta_{n} \triangleq \sqrt{\frac{2\log(1/\epsilon_n)}{n}}C_X(P,Q).
\label{boundlowerR2}
\end{equation}

Finally, replacing $\delta_{n}$ in Eq. (\ref{cover_converse}) and taking logarithm, we have that for any set $\mathcal{B}_n$
satisfying the assumptions of Lemma \ref{lemma_needed}:
\begin{equation}
-\frac{1}{n}\log(Q^n(\mathcal{B}_n))  \leq D(P\|Q)+\frac{\log\left (\frac{1}{1-\epsilon_n-\delta_{n}}\right )}{n}+\delta_n.
\end{equation}
%Since $\mathcal{B}_n$ is arbitrary,  
Therefore, we can choose the optimum set which implies that
\begin{equation}
-\frac{1}{n}\log(\beta_n(\epsilon_n))  \leq D(P\|Q)+\frac{\log\left (\frac{1}{1-\epsilon_n-\delta_{n}}\right )}{n}+\delta_{n}.
\end{equation}
This concludes the proof.
%\end{proof}

\ifCLASSOPTIONcaptionsoff
  \newpage
\fi



\bibliographystyle{IEEEtran}
\bibliography{referencias}
\end{document}